\documentclass[aps,superscriptaddress,amsmath,amssymb,twocolumn,amsfonts,floatfix, longbibliography]{revtex4-2}
\usepackage{mathtools}
\usepackage{braket}
\usepackage[dvipsnames]{xcolor}
\usepackage{float}
\usepackage{subfigure}
\usepackage{upgreek}
\usepackage{pgffor}
\usepackage{mathrsfs}
\usepackage{float}
\usepackage{silence}
\usepackage[colorlinks=true,linktoc=page,linkcolor=blue,citecolor=magenta,urlcolor=Fuchsia]{hyperref}
\usepackage[normalem]{ulem}
\usepackage{sidecap,tikz}
\usepackage{orcidlink}

\newcommand{\beq}{\begin{equation}}
\newcommand{\eeq}{\end{equation}}
\newcommand{\bea}{\begin{eqnarray}}
\newcommand{\eea}{\end{eqnarray}}

\newcommand{\eqn}[1] {Eq.~(\ref{#1})}
\newcommand{\fig}[1]{Fig.~\ref{#1}}
\newcommand{\tab}[1]{Table~\ref{#1}}

\mathchardef\mhyphen="2D 

\newcommand{\non}{\nonumber}

\begin{document}

\title{Zeeman Quantum Geometry as a Probe of Unconventional Magnetism}

\author{{Neelanjan Chakraborti}}
\email{neelanjanc23@iitk.ac.in}
\affiliation{Department of Physics, Indian Institute of Technology, Kanpur 208016, India}
\author{{Sudeep Kumar Ghosh}\,\orcidlink{0000-0002-3646-0629}}
\thanks{Jointly supervised this work}
\email{skghosh@iitk.ac.in}
\affiliation{Department of Physics, Indian Institute of Technology, Kanpur 208016, India}
\author{{Snehasish Nandy}}
\thanks{Jointly supervised this work}
\email{snehasish@phy.nits.ac.in}
\affiliation{Department of Physics, National Institute of Technology Silchar, Assam 788010, India}

\begin{abstract}
Unconventional magnets with momentum-dependent spin-splitting but zero net magnetization form a recently identified class of collinear magnets that are challenging to probe via conventional means. We show that these systems can be distinguished through their intrinsic gyrotropic magnetic (IGM) currents, enabled by the Zeeman quantum geometry, which captures the coupled response of electronic states to momentum translation and spin rotation. Examining two prototypical two-dimensional unconventional magnets with Rashba spin–orbit coupling, a time-reversal-broken $d$-wave altermagnet and a time-reversal-symmetric $p$-wave magnet, we uncover a direct link between crystalline symmetry, spin-split band structures, and transport signatures. The $d_{x^2-y^2}$-wave altermagnet exhibits both transverse conduction and longitudinal displacement IGM currents, whereas the $p$-wave magnet supports only a transverse conduction IGM current. Remarkably, the mixed $d$-wave altermagnet supports all four types of IGM currents, including a longitudinal conduction current enabled by symmetric (Zeeman) Berry curvature that is forbidden in conventional quantum geometry. These responses, measurable via Hall transport and optical probes, persist even when conventional quantum geometry-driven linear responses vanish, offering unique access to hidden spin-split band structures. Our results establish Zeeman quantum geometry as both a diagnostic tool and a design principle for novel magnetic materials.

\end{abstract}

\maketitle

\section{Introduction}
A recently discovered class of collinear magnets, collectively termed as unconventional magnets, exhibit the unusual coexistence of momentum-dependent spin splitting and zero net magnetization~\cite{Kusunose_2019, Zunger_2020, Libor_2021, Jungwirth_2022, Tomas_2022}. In these systems, spin polarization originates from non-relativistic effects enforced by crystalline symmetries, allowing spin splitting even in the absence of spin–orbit coupling (SOC). Two broad subclasses arise depending on the parity of the magnetic order. Even-parity orders, such as $d$-wave altermagnets, break time-reversal symmetry (TRS) while preserving inversion symmetry~\cite{Jungwirth_2022, Tomas_2022, Igor_2022, Sinova_2022, Spaldin_2024}. Odd-parity orders, exemplified by unconventional $p$-wave magnets, preserve TRS but break inversion symmetry~\cite{Libor_2023,Linder_2024_PRL}. Their vanishing net magnetization distinguishes them from both ferromagnets and conventional antiferromagnets, yet simultaneously hinders their detection with standard magnetic probes. Nonetheless, their symmetry-protected spin-split band structures enable unconventional responses, with far-reaching implications for spintronics, quantum information, and correlated electron systems~\cite{Yao_2024,Lorenzo2025,Kriegner_2023,Libor_2022,Nandy_2025,Hariki_2024, Antonenko_2024, Farajollahpour_2025,Jungwirth2025,Song_2025,Lin_2025}.

The quantum geometric tensor (QGT)~\cite{Xiao_2010} characterizes the geometry of Bloch states in Hilbert space and has been directly linked to unconventional Hall and longitudinal transport responses in unconventional magnets~\cite{Cayao_2025,Liu_2025_Nature,Kamra_2024_Nature,Ezawa2024,Ezawa2025,Yamada2025,gorsai_2024_PRL}. In two-dimensional (2D) $d$-wave altermagnets, the combined symmetry of $\hat{C}_{4z}$ and TRS suppresses lower-order contributions, yielding a dominant third-order anomalous Hall effect governed by quantum geometry~\cite{Fang_2024}. Additional mechanisms for anomalous transport have been proposed, including wave-packet magnetic moments~\cite{Ganesh_2025}, light–strain coupling~\cite{Farajollahpour_2025, Keigo_2025}, domain-wall orbital magnetization~\cite{Sorn_2025}, and external magnetic fields~\cite{Rao_2024, Rafael_2025, Tewari_2025,Xu_2020, Furukawa2017, Ni2021, Bing_2020, Ni2021}. Extending beyond the conventional formulation, the Zeeman quantum geometric tensor (ZQGT)~\cite{Xiang_2025_PRL} incorporates both momentum translations and spin rotations of Bloch states, giving rise to new transport phenomena in spin–orbit–coupled systems~\cite{chakraborti2025_TIAM, chakraborti2026PST, chakraborti2026giantspinmagnetization,Cao_2025,Xiang_2025_spin,Furukawa2017}. Despite its natural relevance to symmetry-protected spin-split bands, the role of ZQGT in unconventional magnets remains unexplored.

\begin{figure}[!b]
\centering
\includegraphics[width=\columnwidth]{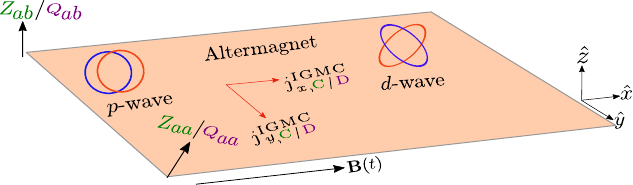}
\caption{\textbf{Schematic of Zeeman quantum geometry-driven linear response:}  
The interplay between the Zeeman Berry curvature ($Z$) and the Zeeman quantum metric ($Q$) gives rise to both longitudinal and transverse (Hall-like) intrinsic gyrotropic magnetic currents (IGMCs) in the linear response of unconventional magnets. Unlike conventional quantum geometry, which forbids such responses, the symmetric part of $Z$ and the antisymmetric part of $Q$ enable both conduction (C) and displacement (D) IGMCs, generating current components parallel as well as perpendicular to the applied oscillating magnetic field. Here, $a$ and $b$ denote spatial indices taking values in ${x,y}$.
}
\label{fig:schematic}
\end{figure}

This motivates the central question of this work: Can Zeeman quantum geometry provide direct and experimentally accessible signatures that distinguish unconventional magnets? To address this, we analyze \emph{linear} transport response arising from the ZQGT, schematically illustrated in \fig{fig:schematic}, in two representative 2D unconventional magnets: a $d$-wave altermagnet and a $p$-wave magnet. We derive analytical expressions for the Zeeman Berry curvature (ZBC) and Zeeman quantum metric (ZQM), and calculate Zeeman quantum geometry-driven intrinsic gyrotropic magnetic (IGM) currents under oscillating magnetic fields. The resulting symmetry-governed IGM conductivities, arising from the interplay between the symmetric and anti-symmetric parts of the ZBC and ZQM, establish Zeeman quantum geometry as a robust diagnostic of spin-split band structures in unconventional magnets. 

\begin{figure*}[!t]
\centering
\includegraphics[width=\textwidth]{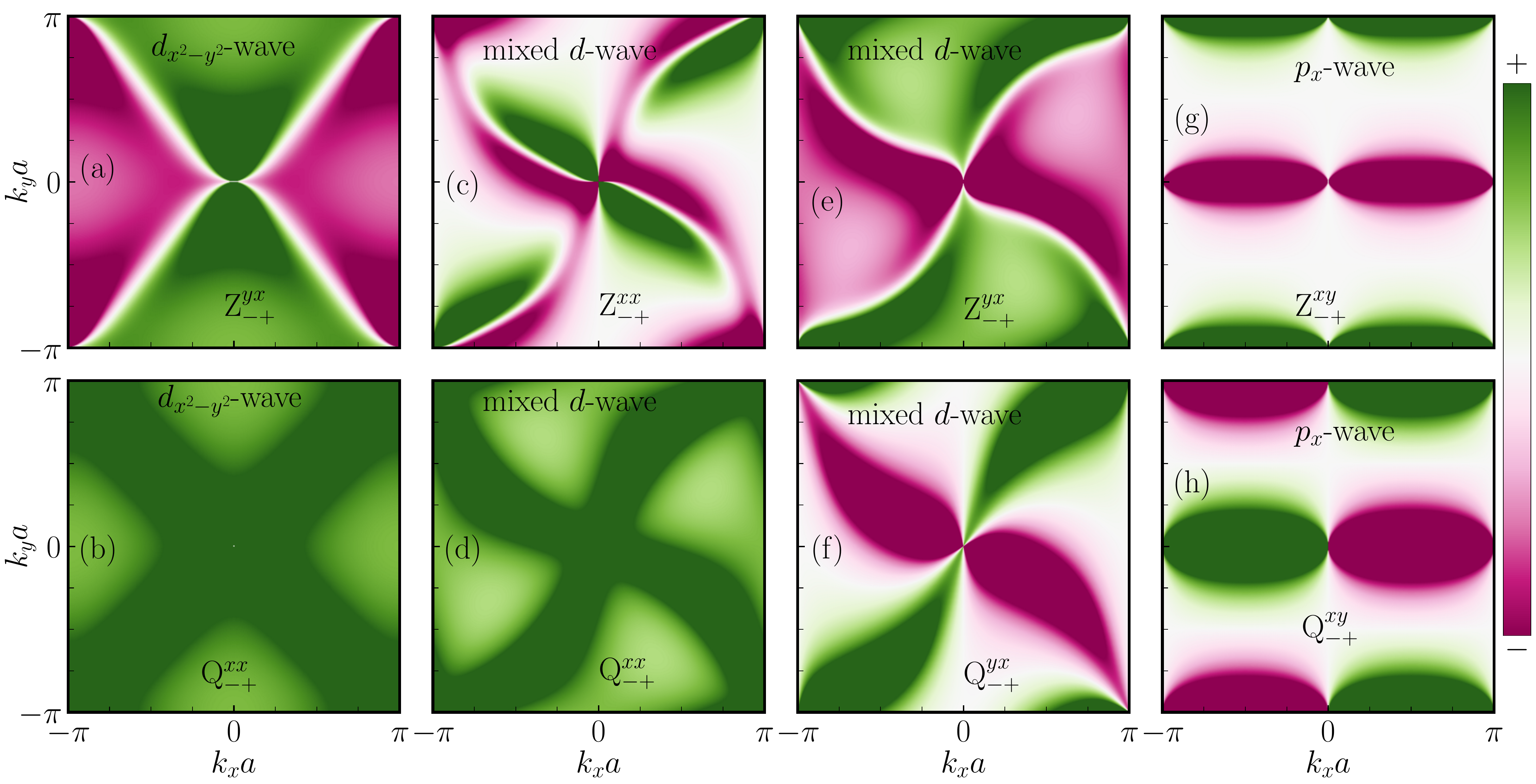}
\caption{\textbf{Zeeman quantum geometry in unconventional magnets:} Distribution of different components of the ZBC ($Z$) and ZQM ($Q$) obtained from the lattice model of the three different types of unconventional magnets. (a) and (b): $Z_{- +}^{yx}$ and $Q_{- +}^{xx}$  are shown for the $d_{x^2 - y^2}$-wave altermagnet. (c)--(f): $Z_{- +}^{xx}$, $Q_{- +}^{xx}$, $Z_{- +}^{yx}$, and $Q_{- +}^{yx}$ are shown for the mixed $d$-wave altermagnet. (g) and (h): $Z_{- +}^{xy}$ and $Q_{- +}^{xy}$ are shown for the $p_x$-wave unconventional magnet. Parameters used: $t = 0.35t'$, $\lambda = 0.2t'$, $t_{am} = 0.5t'$, $t_{am}^\prime = 0.2t'$, $T = 1 \, \text{K}$ and $t' = 1 \, \text{eV}$.}
\label{Geometric_Quantities}
\end{figure*} 


\section{Zeeman quantum geometry in unconventional magnets} The QGT decomposes into a real part, the quantum metric, and an imaginary part, the Berry curvature~\cite{Xiao_2010}. Conventionally, it is defined from the quantum distance between Bloch states $\ket{u^{\xi}_{m\mathbf{k}}}$ differing by infinitesimal momentum shifts, with $m$, $\mathbf{k}$, and $\xi$ labeling the band, crystal momentum, and spin indices, respectively. A generalized formulation~\cite{Xiang_2025_PRL} incorporates both momentum translations and spin rotations, allowing the quantum distance between neighboring states in Hilbert space to be written in a gauge-invariant form, with gauge invariance ensured by the condition $p\neq m$, as
\begin{align}
    ds^2 &=  \left\lVert {\mathscr R}_{d\theta} {\mathscr T}_{d\mathbf{k}} \ket{ u_{m\mathbf{k}}^\xi} - \ket {u_{m\mathbf{k}}^\xi} \right\rVert^2 \nonumber \\
    &= \sum_{p \neq m} {\mathscr G}_{mp}^{ab} \, dk_a \, dk_b 
    + \frac{1}{4} \sum_{m} {\mathscr S}_{pm}^{ab} \, d\theta_a \, d\theta_b \nonumber \\ 
   &\quad +\frac{1}{2} \sum_{p \neq m} \left({\mathscr Z}_{mp}^{ba} + {\mathscr Z}_{pm}^{ba} \right) \, d\theta_a \, dk_b,
    \label{eq:qgt}
\end{align}
\noindent where $a$ and $b$ denote spatial indices. ${\mathscr T}_{d\mathbf{k}} = e^{-i d\mathbf{k} \cdot \hat{\mathbf{r}}}$ represents an infinitesimal momentum translation generated by the position operator $\hat{\mathbf{r}}$, while ${\mathscr R}_{d\theta} = e^{-i d\theta \cdot \hat{\boldsymbol{\sigma}} / 2}$ represents an infinitesimal spin rotation generated by the spin angular momentum operator $\hat{\boldsymbol{\sigma}}/2$ (with $\hbar = 1$), where $\boldsymbol{\sigma}$ is the vector of Pauli matrices. The conventional QGT, ${\mathscr G}_{mp}^{ab}$, arises solely from momentum translations of Bloch states, while the spin-rotation QGT, ${\mathscr S}_{pm}^{ab}$, arises solely from spin rotations. In contrast, the ZQGT: ${\mathscr Z}_{mp}^{ab} = r_{mp}^a \sigma_{pm}^b = Q_{mp}^{ab} - \frac{i}{2} Z_{mp}^{ab}$, combines both the processes. The ZQM ($Q_{mp}^{ab}$) and the ZBC ($Z_{mp}^{ab}$) are defined as~\cite{Xiang_2025_PRL}
\begin{align}
Q_{mp}^{ab} &= \frac{1}{2} \left( r_{mp}^a \sigma_{pm}^b + r_{pm}^a \sigma_{mp}^b \right),\nonumber\\
Z_{mp}^{ab} &= i \left( r_{mp}^a \sigma_{pm}^b - r_{pm}^a \sigma_{mp}^b \right)\label{eq:ZQGT}.
\end{align}
While the intrinsic linear response from the conventional QGT vanishes in $d$-wave altermagnets~\cite{gorsai_2024_PRL}, the spin-rotation QGT contributes only at second order in external perturbations. This implies that, in unconventional magnets, the ZQGT exclusively determines the intrinsic linear response. 

The ZQGT exhibits distinct symmetry behavior compared to the conventional QGT~\cite{Xiang_2025_PRL}. The ZBC $Z^{ab}_{mp}$ is even under TRS ($\hat{\mathcal{T}}$), while the ZQM $Q^{ab}_{mp}$ is odd, arising from the transformation properties of position and spin matrix elements: $\hat{\mathcal{T}} r^{a}_{mp} = r^{a}_{pm}$ and $\hat{\mathcal{T}}\sigma^{a}_{mp} = -\sigma^{a}_{pm}$. Under inversion symmetry ($\hat{\mathcal{P}}$), however, both $Q^{ab}_{mp}$ and $Z^{ab}_{mp}$ are odd, since $\hat{\mathcal{P}}r^{a}_{mp} = -r^{a}_{mp}$ and $\hat{\mathcal{P}}\sigma^{a}_{mp} = \sigma^{a}_{mp}$. The ZBC and ZQM can be decomposed into symmetric ($S$) and antisymmetric ($A$) components, $\mathscr{Y}_{nm}^{S/A;ab} = \left(\mathscr{Y}_{nm}^{ab} \pm \mathscr{Y}_{nm}^{ba}\right)/2$ with $\mathscr{Y}$ representing either $Q$ or $Z$. In conventional quantum geometry, the Berry curvature is antisymmetric and the quantum metric is symmetric~\cite{Xiao_2010}. In stark contrast, Zeeman quantum geometry allows the ZBC and ZQM to possess both symmetric and antisymmetric components~\cite{jiang2025}.
\begin{table}[t]
\centering
\renewcommand{\arraystretch}{1.5}
\setlength{\tabcolsep}{12pt}
\begin{tabular}{|c|c|c|c|c|}
\hline
Symmetry
&
$\mathcal{G}_{mp}^{\alpha\beta}$
&
$\Omega_{mp}^{\alpha\beta}$
&
$Q_{mp}^{\alpha\beta}$
&
$Z_{mp}^{\alpha\beta}$
\\
\hline
$\mathcal{P}$   & $+$ & $+$ & $-$ & $-$ \\
$\mathcal{T}$   & $+$ & $-$ & $-$ & $+$ \\
$\mathcal{PT}$  & $+$ & $-$ & $+$ & $-$ \\
\hline
\end{tabular}
\caption{\textbf{Symmetry classification of quantum geometric quantities:}
Transformation properties of the conventional quantum metric ($\mathcal{G}_{mp}^{\alpha\beta}$), conventional Berry curvature ($\Omega_{mp}^{\alpha\beta}$), Zeeman quantum metric ($Q_{mp}^{\alpha\beta}$), and Zeeman Berry curvature ($Z_{mp}^{\alpha\beta}$) under parity ($\hat{\mathcal{P}}$), time-reversal ($\hat{\mathcal{T}}$), and combined $\hat{\mathcal{PT}}$ symmetries.}
\label{tab:CQGT_vs_ZQGT}
\end{table}
From the symmetry analysis given in~\tab{tab:CQGT_vs_ZQGT}, it is clear that the CQGT and ZQGT are fundamentally distinct. In particular, the ZQGT exhibits unique symmetry properties absent in the conventional QGT, highlighting its novel nature and its potential to generate unconventional transport contributions. 

To investigate the transport signatures arising due to ZQGT in 2D unconventional magnets, we consider a minimal model Hamiltonian for a single orbital on a square lattice~\cite{Farajollahpour_2025,gorsai_2024_PRL,Roig_2024}:
\bea
\!\!\!\!\! &\mathcal{H}& = \mathcal{H}_0 +  g_{\mathbf{k}}\sigma_z\;,\non\\
\!\!\!\!\! &\mathcal{H}_0& = -2t [\cos(k_{x}) + \cos(k_{y})] 
   + \lambda [\sin(k_{y}) \sigma_{x} - \sin(k_{x}) \sigma_{y}].\non\\ 
\!\!\!\!\!\!\!\label{eq:general_ham}
\eea
Here, $t$ is the nearest-neighbor hopping amplitude, $\lambda$ is the strength of the Rashba SOC and the lattice constant is taken to be unity without any loss of generality. The momentum-dependent form factor $g_{\mathbf{k}}$ characterizes the nature of the unconventional magnetic order and gives rise to the momentum-dependent spin splitting that distinguishes different types of unconventional magnetic orders.
To obtain analytical expressions of ZQGT, we also consider the low energy limit of $\mathcal{H}$ around the $\Gamma$-point that can be written as
\beq
\mathcal{H}' = t (k^2_x + k^2_y - 4) 
   + \lambda (k_y \sigma_{x} - k_x \sigma_{y}) +  g'_{\mathbf{k}}\sigma_z \;,
   \label{eqn:general_low_ham}
\eeq
where, $g'_{\mathbf{k}} = \lim_{\mathbf{k} \to 0} g_{\mathbf{k}}$. The 2D unconventional magnetic systems described by the Hamiltonians $\mathcal{H}$ and $\mathcal{H}'$ have two energy bands with dispersions $\epsilon_{\pm}(\mathbf{k})$ where $+$ ($-$) labels the conduction (valence) band. Since the first term in $\mathcal{H}_0$ [Eq.~(\ref{eq:general_ham})] is proportional to the identity, it does not contribute to either the Berry curvature or the quantum metric.

\begin{table}[!ht]
\centering
\renewcommand{\arraystretch}{2.2}
\setlength{\tabcolsep}{2pt}
\caption{Comparison between the different components of the ZBC ($Z$) and ZQM ($Q$) for the two types of unconventional magnets calculated using the respective low energy model Hamiltonian (\eqn{eqn:general_low_ham}). Here, 
$l_k = \big[t_{am}(k_x^2 - k_y^{2}) + 2t_{am}'k_x k_y\big]$, \,
$m_k = \left(t_p^2 k_x^{2} + \lambda^2 k^2\right)$, \,
$n_k = \left[l_k^2 + \lambda^2 k^2\right]$, \,
$p_k = \big(2t_{am}k_x + 2t_{am}'k_y\big)$, \,
$q_k = \big(-2t_{am}k_y + 2t_{am}'k_x\big)$. Putting $t_{am}' = 0$ corresponds to the $d_{x^2-y^2}$-wave case.
}
\label{tab:ZQGT}
\begin{tabular}{|c|c|c|}
\hline
Quantity & mixed $d$-wave altermagnet & $p$-wave magnet \\[1pt]
\hline
$Z_{\pm \mp}^{yx}$ & 
$\pm \dfrac{\lambda \left(l_k q_k k^2 k_y - l_k^2 k_y^2 - n_k k_x^2\right)}{k^2 n_k^{3/2}}$ &
$\mp \dfrac{\lambda k_x^2 \left(t_p^2 k_y^2 + m_k\right)}{m_k^{3/2}k^2}$ \\[8pt]

\hline
$Z_{\pm \mp}^{xy}$ & 
$\pm \dfrac{\lambda \left(l_k p_k k^2 k_x - l_k^2 k_x^2 - n_k k_y^2\right)}{k^2 n_k^{3/2}}$ &
$\pm \dfrac{\lambda^3 k_y^2}{m_k^{3/2}}$ \\[8pt]

\hline
$Z_{\pm \mp}^{xx}$ & 
$\pm \dfrac{\lambda \left(k_y l_k p_k + \lambda^2 k_x k_y\right)}{n_k^{3/2}}$ &
$\pm \dfrac{\lambda k_x k_y \left(t_p^2 + \lambda^2\right)}{m_k^{3/2}}$ \\[8pt]

\hline
$Z_{\pm \mp}^{yy}$ & 
$\pm \dfrac{\lambda \left(k_x l_k q_k + \lambda^2 k_x k_y\right)}{n_k^{3/2}}$ &
$\mp \dfrac{2\lambda^3 k_x k_y}{m_k^{3/2}}$ \\[8pt]

\hline
$Q_{\pm \mp}^{xx}$ & 
$\pm \dfrac{\lambda (p_k k_x - l_k)}{n_k}$ &
$0$ \\[3pt]

\hline
$Q_{\pm \mp}^{yy}$ & 
$\pm \dfrac{\lambda (q_k k_y - l_k)}{n_k}$ &
$\mp \dfrac{\lambda t_p k_x}{m_k}$ \\[3pt]

\hline
$Q_{\pm \mp}^{xy}$ & 
$\pm \dfrac{\lambda k_y p_k}{n_k}$ &
$\pm \dfrac{\lambda t_p k_y}{m_k}$ \\[3pt]

\hline
$Q_{\pm \mp}^{yx}$ & 
$\pm \dfrac{\lambda k_x q_k}{n_k}$ &
$0$ \\[3pt]
\hline
\end{tabular}
\end{table}

\subsection{$d$-wave altermagnet}
To describe a generic $d$-wave momentum-dependent spin-splitting texture on a square lattice, we consider the phenomenological form factor~\cite{Farajollahpour_2025,fukaya2025}  
\beq \label{eqn:d-wave_form_factor}
g_{\mathbf{k}} = t_{\text{am}}(\cos k_{x} - \cos k_{y}) + 2t_{\text{am}}^{\prime} \sin k_{x} \sin k_{y},
\eeq
where the first and second terms correspond to the leading $d_{x^2-y^2}$ and $d_{xy}$ harmonics with amplitudes $t_{\rm am}$ and $t_{\rm am}'$, respectively. The inclusion of both harmonics provides a natural interpolation between the pure $d_{x^2-y^2}$ and $d_{xy}$ limits, allowing for a continuous evolution of the nodal structure and the momentum-space anisotropy of the spin-splitting texture in this mixed $d$-wave altermagnet. Microscopically, the $d_{x^2-y^2}$ and $d_{xy}$ harmonics are naturally associated with nearest- and second-nearest-neighbor hopping processes on the square lattice, respectively~\cite{Farajollahpour_2025}. In the presence of Rashba SOC, the $d$-wave altermagnet Hamiltonian breaks $\hat{\mathcal{P}}$, $\hat{\mathcal{T}}$, and combined $\hat{\mathcal{P}}\hat{\mathcal{T}}$ symmetries.

The ZBC and the ZQM are computed: numerically for the lattice model Hamiltonian (\eqn{eq:general_ham}) with representative results shown in \fig{Geometric_Quantities}, and analytically for the low-energy model (\eqn{eqn:general_low_ham}) as summarized in \tab{tab:ZQGT}. For the $d_{x^2 - y^2}$-wave case ($t_{\text{am}}' = 0$ in \eqn{eqn:d-wave_form_factor}), the diagonal ($xx$ and $yy$) components of the ZQM are purely symmetric, while the off-diagonal ($xy$ and $yx$) components are purely antisymmetric. In contrast, the off-diagonal ${yx}$-component of the ZBC contains both symmetric and antisymmetric contributions. The antisymmetric part $Z^{A;yx}$ forms a monopole-like structure~$(\propto k^2)$ around the $\Gamma$ point, while the symmetric part $Z^{S;yx}$ exhibits a quadrupolar structure $(\propto k_x k_x)$. Moreover, since $Z^{xx}$ and $Q^{yx}$ are proportional to $k_x k_y$, they change sign under a $90^\circ$ rotation in momentum space, characteristic of a quadrupole, while $Q^{xx}$ remains invariant, showing monopole-like behavior. Importantly, the quantum metric scales linearly with the altermagnetic strength $t_{\text{am}}$ and vanishes in its absence, whereas the Berry curvature remains finite even without altermagnetic order. These features are consistent between the lattice model (\fig{Geometric_Quantities}(a) and (b)) and the low energy model results.

When both $t_{\text{am}}$ and $t_{\text{am}}'$ are finite, the additional $d_{xy}$-wave component modifies the ZBC and ZQM into twisted multipole structures, as shown in \fig{Geometric_Quantities}(c)--(f). In this case, $Z^{xx}$ and $Q^{yx}$ are no longer strictly proportional to $k_x k_y$, indicating a deviation from perfect quadrupolar symmetry. Instead, $Z^{yx}$ and $Q^{xx}$ exhibit twisted patterns while preserving their signs near the $\Gamma$ point. A closer inspection reveals that $Z^{yx}$ contains both antisymmetric and symmetric components: the antisymmetric part develops a twisted monopole centered at $\Gamma$, whereas the symmetric part forms a quadrupolar structure. By contrast, $Z^{xx}$ and $Q^{xx}$ lack antisymmetric parts-- the symmetric $Z^{xx}$ forms a twisted quadrupolar Berry curvature, while $Q^{xx}$ acquires a twisted monopole-like structure. Meanwhile, $Q^{yx}$ contains both symmetric and antisymmetric parts: the antisymmetric contribution forms a quadrupole, while the symmetric contribution generates a twisted monopole absent in the pure $d_{x^2-y^2}$-wave case. These behaviors, obtained from the lattice-model calculations (\fig{Geometric_Quantities}(c)--(f)), are in good agreement with that of the low-energy model (\tab{tab:ZQGT}).

\subsection{Unconventional $p$-wave magnet}
An unconventional $p$-wave magnetic state is characterized by an odd-parity momentum-dependent spin splitting described by the form factor $g_{\mathbf{k}}$, which breaks inversion symmetry while remaining invariant under time-reversal operation. For a $p_x$-wave texture,
\[
g_{\mathbf{k}}=t_p\sin k_x,
\]
where $t_p$ denotes the strength of the spin splitting~\cite{fukaya2025}. Such odd-parity spin splittings have recently been proposed to arise from several distinct microscopic mechanisms, including noncollinear magnetic textures, hidden orbital ferrimagnetism in collinear magnets, and orbital-current-driven magnetism. Here, we adopt this form as a representative low-energy model and investigate the corresponding quantum geometric and transport responses. Representative numerical results for the ZBC and quantum metric components of the lattice model (\eqn{eq:general_ham}) are shown in \fig{Geometric_Quantities}(f) and (h), with analytical expressions from the low-energy model listed in \tab{tab:ZQGT}. In this case, the diagonal components of the ZBC are symmetric, while the off-diagonal components contain both symmetric and antisymmetric contributions. The antisymmetric parts of $Z^{yx}$ and $Z^{xy}$ form monopolar structures of opposite sign, whereas their symmetric parts display imperfect quadrupolar distributions. By contrast, $Z^{xx}$ and $Z^{yy}$ yield perfect quadrupolar structures, being proportional to $k_xk_y$. For the ZQM, $Q^{xy}$ and $Q^{yy}$ behave as perfect dipoles proportional to $k_x$ and $k_y$, respectively, and therefore change sign around the $\Gamma$ point. Notably, $Q^{xx}$ and $Q^{yx}$ vanish (and for a $p_y$-wave magnet, $Q^{yy}$ and $Q^{xy}$ vanish), implying that the symmetric and antisymmetric contributions coincide for $Q_{xy}$. Finally, while all components of the ZQM vanish in the absence of $p$-wave magnetic order, the components of the ZBC remain finite.

It is also evident from~\tab{tab:ZQGT} that all nonzero components of the ZQGT are proportional to the Rashba SOC strength $\lambda$. As a result, the ZQGT vanishes identically for $\lambda=0$, establishing that spin mixing induced by Rashba SOC is essential for realizing a finite Zeeman quantum geometry. In the two-dimensional unconventional magnetic systems considered here, Rashba SOC is expected to arise naturally from inversion-symmetry breaking associated with surfaces, interfaces, substrates, external electric fields, or intrinsically non-centrosymmetric crystal structures~\cite{Amundsen_2024,Kapri_2025}.

\begin{figure}[!b]
\centering
\includegraphics[width=\columnwidth]{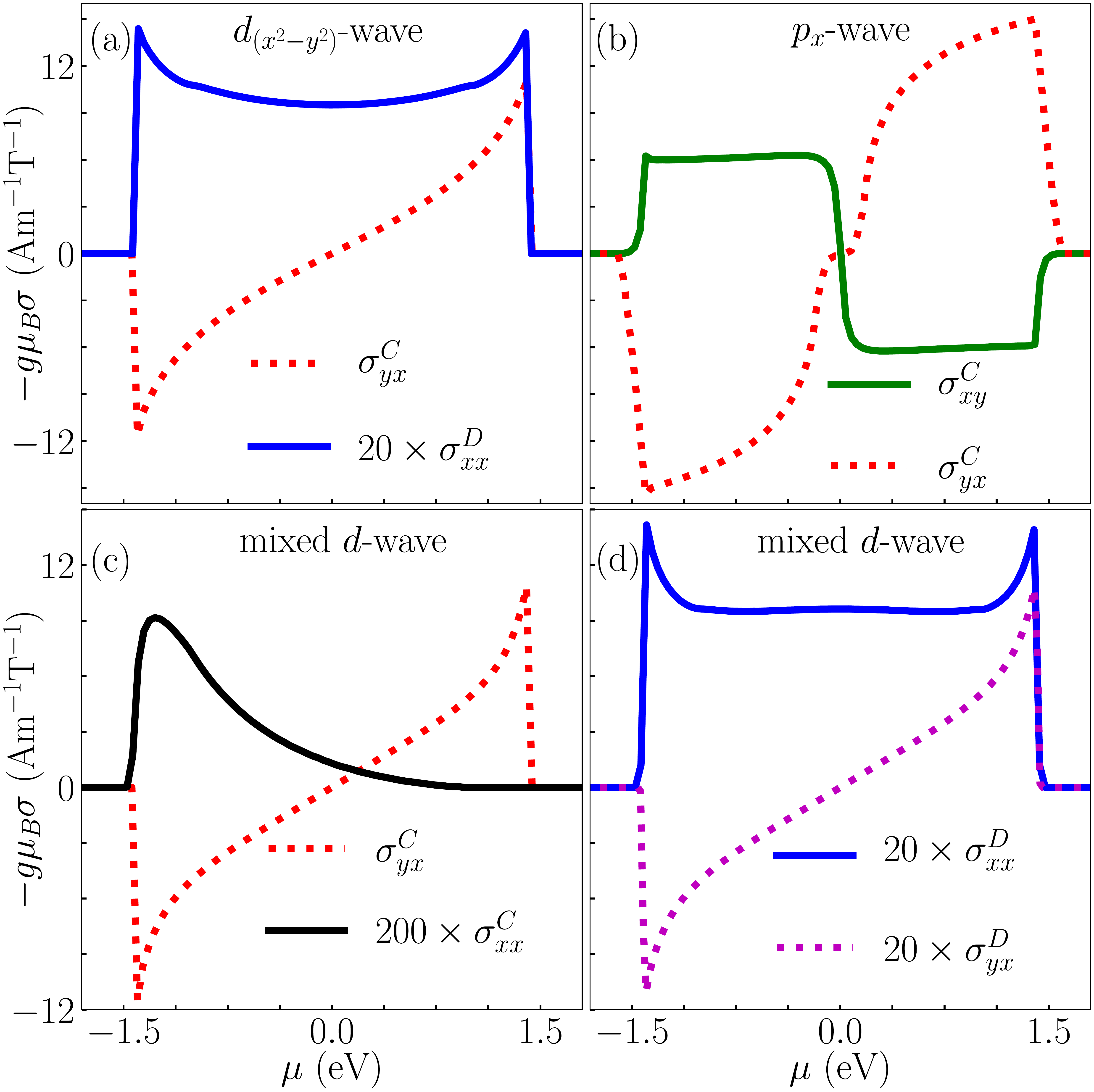}
\caption{\textbf{Zeeman quantum geometry-induced intrinsic gyrotropic magnetic (IGM) conductivities:} (a) In the $d_{x^2-y^2}$-wave altermagnet, a transverse (Hall-like) conduction IGM conductivity ($\sigma_{yx}^C = - \sigma_{xy}^C$) and a longitudinal displacement IGM conductivity ($\sigma_{xx}^D = -\sigma_{yy}^D$) appear. (b) In the $p$-wave magnet, only transverse conduction IGM conductivities ($\sigma_{xy}^C \neq \sigma_{yx}^C$) are present, with no displacement response. (c,d) In the mixed $d$-wave altermagnet, all four conductivity components emerge, reflecting the symmetry-mixed structure. The responses are computed using $\omega = 10^{12}\,\text{Hz}$ and other parameters are the same as in \fig{Geometric_Quantities}.}
\label{conductivity}
\end{figure}

\begin{table*}
\centering
\caption{\textbf{Summary of symmetry properties of ZQGT and corresponding IGM conductivities for different 2D unconventional magnets:} The $d_{x^2 - y^2}$-wave and mixed $d$-wave altermagnets break inversion ($\hat{\mathcal{P}}$), time-reversal ($\hat{\mathcal{T}}$), and combined ${\hat{\mathcal{P}}\hat{\mathcal{T}}}$ symmetry, whereas the $p$-wave ordering preserves $\hat{\mathcal{T}}$ symmetry but breaks $\hat{\mathcal{P}}$ symmetry. The presence or absence of the ZQM ($Q^{ab}$), ZBC ($Z^{ab}$), conduction IGM conductivity ($\sigma_{ab}^C$), and displacement IGM conductivity ($\sigma_{ab}^D$) are listed. A cross (\text{\sffamily X}) indicates a vanishing contribution due to symmetry constraints; a tick (\checkmark) indicates allowed contributions.}
\label{tab:symmetry}
\renewcommand{\arraystretch}{0.95}
\setlength{\tabcolsep}{5pt}
\begin{tabular}{|c|c|c|c|c|c|}
\hline
\begin{tabular}{@{}c@{}}Magnetic \\ Order \end{tabular} & \begin{tabular}{@{}c@{}}Symmetry \\ ($\hat{\mathcal{P}}$, $\hat{\mathcal{T}}$, $\hat{\mathcal{PT}}$)\end{tabular} & $Q^{ab}$ & $Z^{ab}$ & $\sigma_{ab}^{C}$ & $\sigma_{ab}^{D}$ \\ 
\hline
$d_{x^2 - y^2}$-wave &  ($\text{\sffamily X},\ \text{\sffamily X},\ \text{\sffamily X}$) & $Q^{yx},Q^{xy},Q^{xx},Q^{yy}$ &$Z^{yx},Z^{xy},Z^{xx},Z^{yy}$ & $\sigma_{xy}^{C}=-\sigma_{yx}^{C}$ & $\sigma_{xx}^{D}=-\sigma_{yy}^{D}$\\ 
$p_{x}$-wave & ($\text{\sffamily X},\ \checkmark,\ \text{\sffamily X}$) &  $Q^{xy},Q^{yy}$ & $Z^{yx},Z^{xy},Z^{xx},Z^{yy}$ & $\sigma_{xy}^{C},\sigma_{yx}^{C}$ &  $\text{\sffamily X}$\\
Mixed $d$-wave &  ($\text{\sffamily X},\ \text{\sffamily X},\ \text{\sffamily X}$) &$Q^{yx},Q^{xy},Q^{xx},Q^{yy}$ & $Z^{yx},Z^{xy},Z^{xx},Z^{yy}$ & $\sigma_{xy}^{C}=-\sigma_{yx}^{C},\ \sigma_{xx}^{C}=\sigma_{yy}^{C}$ & $\sigma_{xx}^{D}=-\sigma_{yy}^{D},\ \sigma_{xy}^{D}=\sigma_{yx}^{D}$ \\ 
\hline
\end{tabular}
\end{table*}

\section{Intrinsic gyrotropic transport in unconventional magnets}
An oscillating magnetic field $\mathbf{B}(t) = \mathbf{B}_0 \cos(\omega t)$ with frequency $\omega$, gives rise to two types of IGM conductivities: the conduction response $\sigma^{C}_{ab}$ and the displacement response $\sigma^{D}_{ab}$. These are governed by the ZBC and ZQM, and are computed upto the linear order in $\omega$ as~\cite{Xiang_2025_PRL}:
\begin{equation}
\!\!\!\sigma^C_{ab} =\sum_{m p} \int_k f_m Z^{ab}_{mp} \;\;\;\; \&
\;\;\;\; \sigma^D_{ab} =\sum_{m p} \int_k f_m \frac{(2 \omega)}{\epsilon_{pm}} Q^{ab}_{mp},
\label{eq:Current}
\end{equation}
where $f_m$ is the Fermi-Dirac distribution and $\epsilon_{pm}$ is the energy difference between the $p$-th and $m$-th bands. Here, we consider frequencies of the oscillating magnetic field well below the threshold for interband absorption, i.e. $\hbar \omega \ll \epsilon_{pm}$. We note that  the conduction IGM conductivity is a fermi surface quantity that follows the relation $\sum_p Z_{mp}^{ab} = \partial_a \sigma^b_{m}$, whereas the displacement IGM conductivity is a fermi sea quantity. It is evident that both conductivities originate purely from band geometric quantities and are independent of the relaxation time $\tau$, confirming their intrinsic nature. In contrast, it has been proposed that IGM currents can also arise from the intrinsic magnetic moment (orbital and spin) of Bloch electrons on the Fermi surface~\cite{Zhong_2016}. However, this contribution is relaxation-time dependent and can thus be distinguished from the ZQGT-driven contribution.

For the $d_{x^2-y^2}$-wave altermagnet, $Q^{yx}$ and $Z^{xx}$ do not contribute to the displacement or conduction IGM currents because of their perfect quadrupolar symmetry. As seen from Eq.~(\ref{eq:Current}), integration over the Brillouin zone cancels contributions from regions with $k_x k_y > 0$ and $k_x k_y < 0$, yielding zero net response. In contrast, $Q^{xx}$ and $Z^{yx}$ are even in momentum and produce finite contributions: the antisymmetric (monopole-like) component of $Z^{yx}$ drives the transverse conduction IGM conductivity $\sigma_{yx}^C$, while $Q^{xx}$ generates the longitudinal displacement IGM conductivity $\sigma_{xx}^D$. These obey the symmetry relations $\sigma_{xy}^C = -\sigma_{yx}^C$ and $\sigma_{xx}^D = -\sigma_{yy}^D$. As shown in Fig.~\ref{conductivity}(a)--(b), $\sigma_{yx}^C$ changes sign at $\mu = 0$, reflecting the reversal of the ZBC in Fig.~\ref{Geometric_Quantities}(a), whereas $\sigma_{xx}^D$ remains sign-definite since $Q^{xx}$ does not change sign across the Brillouin zone [Fig.~\ref{Geometric_Quantities}(b)]. The emergence of linear Hall and longitudinal IGM conductivities, governed respectively by the antisymmetric ZBC and symmetric ZQM, constitutes a novel feature absent in conventional quantum geometry~\cite{gorsai_2024_PRL} and represents a central result of this work.

When both $d_{x^2-y^2}$-wave and $d_{xy}$-wave components coexist (the mixed $d$-wave altermagnet case), the system supports all IGM currents. As in the pure $d_{x^2-y^2}$-wave case, the antisymmetric ZBC drives the transverse conduction current $\sigma_{yx}^C$ and the symmetric ZQM governs the longitudinal displacement current $\sigma_{xx}^D$. However, two additional contributions arise: (i) a longitudinal conduction current $\sigma_{xx}^C$ [Fig.~\ref{conductivity}(c)], and (ii) a transverse displacement current $\sigma_{xy}^D$ [Fig.~\ref{conductivity}(d)]. These originate from the symmetric parts of $Z^{xx}$ and $Q^{xy}$ respectively, which no longer exhibit perfect quadrupole symmetry in the mixed state. The mixing of $d_{x^2-y^2}$ and $d_{xy}$ components produces a ``twisted'' quadrupole in both $Z^{xx}$ and $Q^{yx}$ [Fig.~\ref{Geometric_Quantities}(c) and (f)], yielding finite contributions upon Brillouin-zone integration. The IGM conductivities in this case satisfy: $\sigma_{xy}^C = -\sigma_{yx}^C$, $\sigma_{xx}^C = \sigma_{yy}^C$, $\sigma_{xy}^D = \sigma_{yx}^D$, and $\sigma_{xx}^D = -\sigma_{yy}^D$.  

For the $p_x$-wave unconventional magnet, only conduction IGM conductivities $\sigma_{xy}^C$ and $\sigma_{yx}^C$ appear, arising from finite $Z^{xy}$ and $Z^{yx}$. TRS ensures that all displacement currents vanish, which is further supported by the dipole-like structure of the quantum metric components, has been discussed in details in Appendix-II. The transverse conduction IGM current is driven by both the symmetric and antisymmetric parts of the ZBC. Notably, $\sigma_{xy}^C$ is determined by the sum: $\left({Z_{nm}^{S;ab} + Z_{nm}^{A;ab}}\right)$, whereas $\sigma_{yx}^C$ is determined by their difference: $\left({Z_{nm}^{S;ab} - Z_{nm}^{A;ab}}\right)$. Fig.~\ref{conductivity}(b) shows the variation of $\sigma_{xy}^C$ and $\sigma_{yx}^C$ with chemical potential. Both conductivities vanish at $\mu = 0$ and exhibit asymmetric behavior, though with unequal magnitudes: $\sigma_{xy}^C \neq -\sigma_{yx}^C$. The sign change in $\sigma_{xy}^{C}$ and $\sigma_{yx}^{C}$ at $\mu = 0$ is consistent with the sign reversal of the ZBC ($Z^{xy}$ and $Z^{yx}$). In the case of a $p_y$-wave magnet, the displacement current vanishes, while both $\sigma_{xy}^C$ and $\sigma_{yx}^C$ can remain nonzero. A summary of the symmetry properties and the corresponding nonzero IGM conductivities for the $d$-wave and $p_x$-wave cases is provided in Table~\ref{tab:symmetry}.

Note that perturbative corrections to both the conventional and Zeeman Berry curvatures arising from the magnetic field are not considered here, as they contribute only to nonlinear IGM currents~\cite{Wang_2024, Wang_2024_PRL}. To estimate the magnitude of the linear IGM current, we take RuO$_2$ as a representative $d_{x^2-y^2}$-wave altermagnet candidate~\cite{Roser_2024_PRB,Liu_2024_PRL,Xiao_2025_NPJ}, with parameters $t_{am} = 1 \;\text{eV}$, $t = 1 \;\text{eV}$, $\lambda = 0.4 \;\text{eV}$, and $\mu = 0.3 \;\text{eV}$~\cite{Sinova_2020_Science}. For a Hall-bar device of lateral dimension $\sim 200 \;\mu\text{m}$ and resistance $\sim 10^3 \;\Omega$, applying a weak magnetic field $B = 20 \;\text{G}$ at a low frequency $\omega \approx 10^2 \;\text{Hz}$ yields a conduction-type Hall IGM voltage of about $2.3 \;\text{mV}$-- readily measurable with standard transport setups~\cite{Nagaosa_2010}. In contrast, detecting the displacement-type IGM current requires terahertz frequencies ($\omega \sim 10^{13} \;\text{Hz} \sim 0.6 \;\text{meV}$), where the resulting voltage of $\sim 0.014 \;\text{mV}$ remains within the reach of modern THz spectroscopy~\cite{Matsuda_2020}. The ZQGT-driven intrinsic conduction and displacement IGM conductivity thus provides a powerful experimental probe to identify different unconventional magnetic order. In particular, although RuO$_2$ has been proposed as a prototypical $d$-wave altermagnet~\cite{Karube_2022, Bai_2022, Liao_2024, Furukawa2017}, recent neutron scattering, $\mu$SR, and spin-ARPES measurements have reported no clear signatures of long-range antiferromagnetic order or an altermagnetic band structure~\cite{Liu_2024, Hiraishi_2024, Kebler_2024}. Observation of a $d_{x^{2}-y^{2}}$-wave altermagnetic state, manifested through the simultaneous presence of a transverse conduction current and a longitudinal displacement current, would therefore help to resolve this outstanding issue.


\section{Summary and conclusions}
Unconventional magnets, characterized by momentum-dependent spin-split band structures but zero net magnetization, are notoriously difficult to characterize using traditional methods. We demonstrate that Zeeman quantum geometry offers a powerful diagnostic by encoding the coupled response of Bloch states to momentum translation and spin rotation. This mechanism generates IGM currents, which serve as distinct linear transport signatures in unconventional magnets with Rashba SOC. By analyzing two prototypical 2D magnets, a d-wave altermagnet and a p-wave magnet, we established direct links between crystalline symmetry, spin-split band structure, and transport. Specifically, the $d_{x^2-y^2}$-wave altermagnet supports both transverse conduction and longitudinal displacement IGM currents, whereas the p-wave magnet hosts only a transverse conduction IGM current. Most strikingly, the mixed $d$-wave altermagnet realizes all four possible IGM currents, including a component enabled by symmetric ZBC absent in conventional quantum geometry. These results demonstrate that the proposed diagnostic signatures are governed by symmetry-enforced patterns of allowed and forbidden IGM conductivity tensor components, which explicitly depend on the form factors of different unconventional magnetic orders and their underlying symmetries, rather than on the specific choice of model parameters. Additionally for $d$- and $p$-wave unconventional magnets, the CQGT contribution is identically zero , whereas the SRQGT contributes only at second order, as discussed in detail in Appendix~I. Therefore, at linear order, the ZQGT-driven response constitutes the sole nonvanishing contribution. These robust responses, which are measurable via Hall transport and optical probes, persist even when conventional Berry-curvature effects vanish, providing unique access to hidden spin textures. Our results establish the ZQGT as both a vital diagnostic tool and a guiding principle for identifying and designing novel unconventional magnetic materials.

For experimental detection, the proposed intrinsic conduction and displacement gyrotropic magnetic currents are driven solely by oscillating magnetic fields and remain robust against moderate disorder. The conduction IGM current can be detected in a Hall-bar geometry through a measurable Hall voltage, while the displacement IGM current may be probed via circular dichroism, providing direct access to hidden spin textures~\cite{HosurQi2015, MaPesin2015, Zhong_2016}. Although an accompanying electric field is inevitably induced by Faraday’s law, its influence on quantum-geometric currents can be suppressed by aligning the polarization and propagation direction of the magnetic field within the plane of the magnet. 

In addition, extrinsic gyrotropic contributions arising from intrinsic magnetic moments (orbital plus spin) of the Bloch electrons on the Fermi surface or disorder effects (side-jump and skew scattering) are proportional to the scattering time and therefore, can be separated through scaling analysis, allowing their separation via disorder tuning~\cite{Zhong_2016, Nandy_2019, Du_2019}. Furthermore, the intrinsic and extrinsic contributions can also be separated via their frequency dependence since the intrinsic displacement current scales as $\sigma^{D}\propto\omega$, while extrinsic contributions typically follow owing their separation via disorder tuning $i\omega\tau/(1-i\omega\tau)$~\cite{Zhong_2016}. Their temperature dependences are also distinct due to the scattering-induced variation of $\tau$ in the extrinsic response. Promising material platforms include thin films and monolayers of RuO$_2$~\cite{Roser_2024_PRB,Liu_2024_PRL,Xiao_2025_NPJ}, CrSb~\cite{Libor_2024_Nature,Peng_2025_PRB,Cong_2024}, MnTe~\cite{Mazin_2023_PRB,Lee_2024_PRL,Amin_2024_Nature,Kriegner_2024_npj}, and MnF$_2$~\cite{Morano_2025}, where Rashba SOC and symmetry conditions conducive to ZQGT-driven transport are naturally realized. The unconventional magnetic orders considered here, and the resulting IGM currents, can arise in a wide range of magnetic point groups~\cite{Cheong2025,mitscherling2026}. Looking ahead, extending this framework to non-collinear or three-dimensional magnets, correlated regimes, and thermal or optical analogues of IGM currents could open pathways to reconfigurable quantum devices that exploit geometric control of transport. Thus, the Zeeman quantum geometric tensor emerges not only as a diagnostic tool but also as a guiding principle for the design of unconventional magnetic materials with tunable and robust transport functionalities.\\


\noindent \textbf{\em Acknowledgments:} N.~C. acknowledges the Council of Scientific and Industrial Research (CSIR), Government of India, for providing the JRF fellowship. S.~K.~G. acknowledges financial support from Anusandhan National Research Foundation (ANRF) erstwhile Science and Engineering Research Board (SERB), Government of India via the Startup Research Grant: SRG/2023/000934. S.~N. also acknowledges financial support from Anusandhan National Research Foundation (ANRF), Government of India via the Prime Minister's Early Career Research Grant: ANRF/ECRG/2024/005947/PMS.\\

\noindent \textbf{\em Author Contributions:} S.~K.~G. and S.~N. conceptualized and supervised the work. N.~C. performed the calculations and prepared the manuscript draft with guidance and input from S.~K.~G. and S.~N. All authors discussed the results, contributed to data interpretation, and reviewed the final version of the manuscript.\\

\noindent \textbf{\em Funding Declaration:} S.~K.~G. and S.~N. acknowledge financial support from Anusandhan National Research Foundation (ANRF) erstwhile Science and Engineering Research Board (SERB), Government of India respectively via the Startup Research Grant: SRG/2023/000934 and the Prime Minister's Early Career Research Grant: ANRF/ECRG/2024/005947/PMS.

\section{Appendix-I}
\label{sec:Appendix-I}
\renewcommand{\theequation}{A\arabic{equation}}
\setcounter{equation}{0}

The conventional QGT contribution to the linear Hall response originates from the conventional Berry curvature (CBC) $\Omega_{\pm \mp}^{ab}$, which is odd under time-reversal symmetry~\cite{Xiao_2010}. The corresponding anomalous Hall conductivity is given by
\begin{equation}
\sigma_{xy}
=
-e^2 \sum_n
\int_{\mathrm{BZ}}
\frac{d^3k}{(2\pi)^3}
\, f_n(\mathbf{k})
\,\Omega_n^{xy}(\mathbf{k}).
\end{equation}
From the low-energy model of the $d$-wave altermagnet in
Eq.~\eqref{eqn:general_low_ham}, the CBC takes the form
$ \Omega_{\pm\mp}^{xy}(\mathbf{k}) \propto \pm \frac{g_k'}{(\lambda^2 k^2 + g_k'^2)^{3/2}} $, where the angular dependence is entirely governed by $g'_k$. Since $g'_k$ possesses a $d$-wave form factor, the CBC exhibits a characteristic quadrupolar symmetry in momentum space, while for a $p$-wave unconventional magnet $\Omega_{\pm\mp}^{xy}(\mathbf{k})$ is zero. Consequently, the Brillouin-zone integral of the linear Hall conductivity vanishes identically.

The pure spin-rotation QGT contribution, involving matrix elements of the form $\sigma^a_{mp}\sigma^b_{pm}$, couples only to spin-space perturbations. Since the external field enters through the Zeeman coupling, this contribution appears only at higher order in perturbation theory and therefore does not contribute to the intrinsic linear response~\cite{chakraborti2026_nonlinear}.

\section{Appendix-II}
\label{sec:Appendix-II}
\renewcommand{\theequation}{B\arabic{equation}}
\setcounter{equation}{0}

We now provide an explicit symmetry argument for the vanishing of the displacement IGM conductivity in the time-reversal symmetric $p$-wave case. The displacement conductivity is given by
\begin{align}
\sigma^{D}_{ab} = \sum_{mp} \int_{\mathbf{k}} \frac{2 \omega f_m}{\epsilon_{pm}} \, Q^{ab}_{mp}(\mathbf{k}),
\end{align}
where $Q^{ab}_{mp}$ is the Zeeman quantum metric (ZQM). Under time-reversal symmetry $\mathcal{T}$, the momentum transforms as $\mathbf{k} \to -\mathbf{k}$, $\mathbf{r} \to \mathbf{r}$ and $\sigma \to -\sigma$, the matrix elements transform as,
\begin{align}
\mathcal{T}: \quad Q^{ab}_{mp}(\mathbf{k}) \to - Q^{ab}_{mp}(-\mathbf{k}),
\end{align}
i.e., the ZQM is odd under time reversal, as discussed in the manuscript. At the same time, the band energies satisfy $\epsilon_m(\mathbf{k}) = \epsilon_m(-\mathbf{k})$, and hence both the Fermi factor $f_m$ and the energy denominator $\epsilon_{pm}$ are even under $\mathbf{k} \to -\mathbf{k}$.

Therefore, the full integrand of $\sigma^D_{ab}$ transforms as
\begin{align}
\mathcal{I}_{ab}(\mathbf{k}) = \frac{f_m}{2\hbar \omega \,\epsilon_{pm}} \, Q^{ab}_{mp}(\mathbf{k})
\;\;\xrightarrow{\mathcal{T}}\;\;
-\mathcal{I}_{ab}(-\mathbf{k}).
\end{align}
This implies that the integrand is an odd function over the Brillouin zone. Since the Brillouin zone is symmetric under $\mathbf{k} \to -\mathbf{k}$, the integral vanishes identically:
\begin{align}
\sigma^D_{ab} = \int_{\mathbf{k}} \mathcal{I}_{ab}(\mathbf{k}) = 0.
\end{align}

\bibliography{ZQGT}
\end{document}